\documentclass[pra,aps,showpacs,twocolumn]{revtex4}

\usepackage{amsfonts}
\usepackage{amsmath}
\usepackage{amssymb}
\usepackage[dvips]{graphicx}

\begin{document}

\title{Selective Control of the Symmetric Dicke Subspace in Trapped Ions}

\date{\today }

\author{C. E. L\'{o}pez$^{1}$, J. C. Retamal$^{1}$ and E. Solano$^{2,3}$}
\affiliation{$^1$Departamento de F\'{\i}sica, Universidad de Santiago de Chile, Casilla
307 Correo 2, Santiago, Chile \\
$^2$Physics Department, ASC, and CeNS,
Ludwig-Maximilians-Universit\"at, Theresienstrasse 37, 80333
Munich,
Germany \\
$^3$Secci\'on F\'{\i}sica, Departamento de Ciencias, Pontificia
Universidad Cat\'olica del Per\'u, Apartado 1761, Lima, Peru}


\begin{abstract}
We propose a method of manipulating selectively the
symmetric Dicke subspace in the internal degrees of freedom of N trapped ions. We show that the direct access to
ionic-motional subspaces, based on a suitable tuning of
motion-dependent AC Stark shifts, induces a two-level dynamics involving previously selected ionic Dicke states. In this manner, it is possible to produce, sequentially and unitarily, ionic Dicke states with increasing excitation number. Moreover, we propose a probabilistic technique to produce directly any ionic Dicke state assuming suitable initial conditions.
\end{abstract}

\maketitle
\section{Introduction}
Multipartite entangled states play a fundamental role in quantum
information, where these states are used for different
applications including the improvement of spectroscopy towards the
Heisenberg limit~\cite{Leibfried04}. In this sense, general
sequential techniques for building entangled multipartite states
have been recently proposed~\cite{Schoen05}.  In
Ref.~\cite{Haffner05}, it is described an experiment where the
robust one-excitation symmetric Dicke states~\cite{Dicke54},
called $W$, of $N \leq 8$ ions are prepared in their electronic
levels with the aid of $N$ entangling pulses. Also, a maximally
entangled (GHZ) state with six ions has been experimentally
realized~\cite{Leibfried05}.  From a theoretical point of view,
adiabatic ground-state transitions were proposed for generating
GHZ states and symmetric Dicke states with $N / 2$ excitations
in $N$ ions~\cite{Fleischhauer03}. More recently, a method for
generating multi-qubit entangled states via global addressing of
an ion chain in the frame of the Tavis-Cummings model has been
discussed~\cite{Retzker06}. A four-qubit $W$ state with two
excitations has already been realized in linear
optics~\cite{Kiesel07}, which may present astonishing multipartite properties~\cite{Kaszlikowski07}, and more general proposals may be
considered~\cite{Thiel07}. It is well established that a physical
system must fulfill several requirements in order to qualify as a
potential candidate for quantum computing
tasks~\cite{DiVincenzo00}. Among them, overcoming decoherence and
scalability considerations may require not only efficient single-
and two-qubit gates but also the availability of collective
multipartite operations in suitable subspaces.

In this letter, we consider a system composed of $N$ trapped ions
addressed collectively by two laser fields in a global Lambda-type
excitation scheme. We will introduce a method for tailoring the
Hilbert space in order to restrict the quantum dynamics to the
symmetric Dicke subspace. As we show below, this method allows a
different and useful way to manipulate selectively the collective
ionic-motional system. In particular, these multipartite selective
interactions will permit the generation of ionic Dicke states with
any number of excitations in a sequential manner or, through a
probabilistic technique, in a single-shot measurement. This method
is based on global selective interactions characterized by a
proper tuning of collective motion-dependent Stark shifts.
Selective interactions with a single atom have been proposed in
the realm of cavity QED~\cite{Santos01} and trapped
ions~\cite{Solano00,Solano05}. Furthermore, it has been
demonstrated that they also allow the generation of arbitrary
harmonic oscillator states~\cite{Santos05} and their measurement
via instantaneous interactions~\cite{Santos07}.

\section{The Model}
\begin{figure}[t]
\includegraphics[width=55mm]{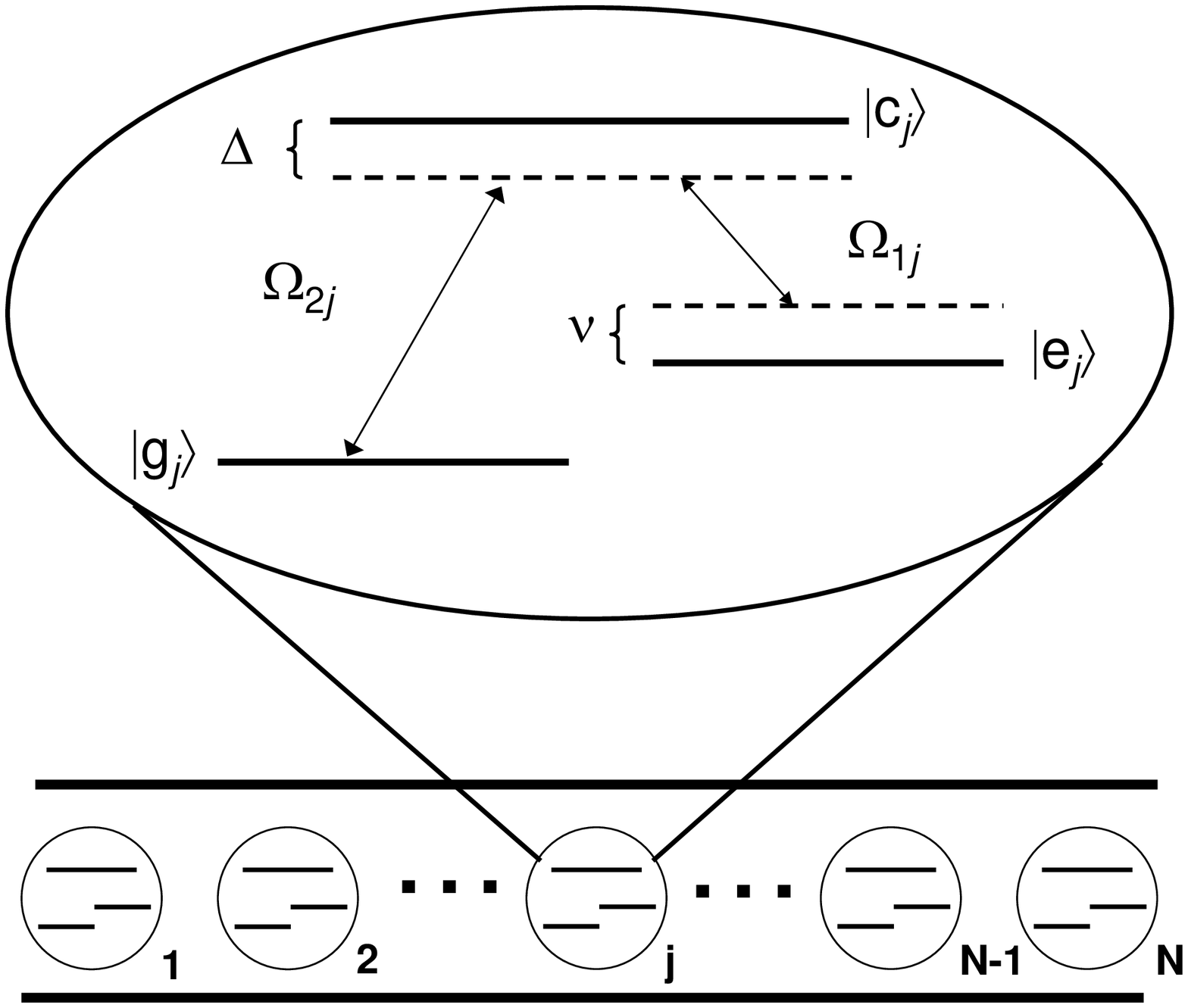}
\caption{N three-level ions in a linear Paul trap where the energy
diagram of the $j$-th ion is displayed.} \label{fig1}
\end{figure}

Let us consider a Raman laser excitation of $N$ three-level
trapped ions as shown in Fig.~\ref{fig1}. We will make use of
these internal levels and the collective center-of-mass motional
mode associated with the frequency $\nu$. A travelling-wave field
excites the transition between the states $|\mathrm{g}_j\rangle
\leftrightarrow |\mathrm{c}_j\rangle$, with coupling strength
$\Omega_{2j} = \Omega_{2j}( \vec{r}_{j})$ and detuning $\Delta$
($\Delta\gg \Omega_{2j}$). Similarly, a standing-wave field
excites off resonantly the transition between the electronic
internal states $|\mathrm{e}_j\rangle \leftrightarrow
|\mathrm{c}_j\rangle$, with position-dependent coupling strength
$\Omega_{1j} = \Omega_{1j}( \vec{r}_{j})$ and  detuning $\Delta +
\nu \gg \Omega_{1j}$. This scenario is described, after a first
optical rotating-wave-approximation (RWA), by the Hamiltonian
\begin{eqnarray}
\hat{H} & = &\hbar \nu \hat{a}^{\dag }\hat{a}+\hbar \omega
_{e}\sum_{j=1}^{N}|\mathrm{e}_{j}\rangle \langle \mathrm{e}_{j}|
+\hbar \omega _{c}\sum_{j=1}^{N}|\mathrm{c}_{j}\rangle \langle
\mathrm{c}_{j}|   \notag \\
&&  \notag \\
&& + \hbar \bigg\lbrack \cos ( k_{1}\hat{z}) e^{i\omega
_{1}t}\sum_{j=1}^{N}\Omega_{1j}|\mathrm{e}_{j}\rangle \langle
\mathrm{c}_{j}|   \notag \\
&&  \notag \\
&&+e^{-i(k_{2}\hat{z}-\omega
_{2}t)}\sum_{j=1}^{N}\Omega_{2j}|\mathrm{g}_{j}\rangle \langle
\mathrm{c}_{j} | +\mathrm{H.c.} \bigg\rbrack.
\end{eqnarray}
We go then to an interaction picture inside the Lamb-Dicke regime: $\eta_i
\sqrt{\bar{n}}\ll 1$, where $\bar n$ is the average phonon number
and $\eta_i \equiv k_i\sqrt{\hbar/2m\nu}$ are the Lamb-Dicke
parameters. In this way, we can adiabatically eliminate levels $|c_j\rangle $,
obtaining the blue-sideband second-order effective Hamiltonian
\begin{equation}
\hat{H}_{\mathrm{eff}}=-\hbar \hat{\Delta} + \hbar ( \hat{a}^{\dag
}\hat{\tilde{J}}^{+} +  \hat{a}\hat{\tilde{J}}^{-} ), \label{HeffN}
\end{equation}
where $\hat{\tilde{J}}^{+} = \sum_{j=1}^{N}\Omega
_{\mathrm{eff}}^j\hat{\sigma} _{j}^{\dag }$, with $\Omega
_{\mathrm{eff}}^j = 2i \eta _{2}\Omega_{1j} \Omega_{2j}^{\ast
} / \Delta$, $\hat{\sigma}^{\dagger}_{j}=|\mathrm{e}_{j}\rangle
\langle \mathrm{g}_{j}|$, and
\begin{eqnarray}
\hat{\Delta} &=&\frac{1}{\Delta }\sum_{j=1}^{N}[1-\eta
_{1}^{2}( 2\hat{a}^{\dag
}\hat{a}+1)]|\Omega_{1j}|^{2}|\mathrm{g}_{j}\rangle
\langle \mathrm{g}_{j}| \notag \\
&&+\frac{1}{\Delta }\sum_{j=1}^{N}|\Omega_{2j}|
^{2}|\mathrm{e}_{j}\rangle \langle \mathrm{e}_{j}|  \label{Delta}
\end{eqnarray}
is the motion-dependent AC Stark shift. In this case, we can
discard terms involving level $|\mathrm{c}_j\rangle$ by assuming
no initial population. The phonon-number dependence of the Stark
shift~(\ref{Delta}) is due to the standing-wave Raman laser, which
together with the travelling wave produce the dynamics of
Eq.~(\ref{HeffN}). Note that AC Stark shifts have already been used for experimental realization of two-qubit gates and multipartite
entanglement~\cite{Schmidt-Kaler04}.

The detuning $\hat{\Delta}$ can be corrected by a fixed position-dependent quantity $\delta_{0}^j$ via DC Stark
shift or retuning of the lasers frequencies. In this manner,
Hamiltonian~(\ref{HeffN}) can be written as
\begin{eqnarray}
\hat{H}_{\mathrm{eff}} & = & - \hbar \sum_{j=1}^{N}\Omega _0^j( \hat{n}-\delta_{0}^j)
|\mathrm{g}_{j}\rangle
\langle \mathrm{g}_{j}|  \notag \\
&& + \hbar ( \hat{a}^{\dag }\hat{\tilde{J}}^{+}+\hat{a}\hat{\tilde{J}}^{-} ),
\label{HeffN2}
\end{eqnarray}
where $\Omega_0^j=2\eta^2_1|\Omega_{1j}|^2/\Delta$. It will be convenient to rewrite the Hamiltonian of Eq.~(\ref{HeffN2}) in the interaction picture with respect to the first term, where it reads
\begin{equation}
\hat{H}_{\mathrm{eff}}^{\mathrm{I}} = \hbar \sum_{j=1}^N \Omega_{\mathrm{eff}}^j
\hat{a}^{\dag} \hat{\sigma}_j^{\dag} e^{-i[\Omega_0^j
(\hat{n} - \delta_0^j) + \sum_{k \ne j} \Omega_0^k |\mathrm{g}_k\rangle\langle
\mathrm{g}_k| ] t} + \mathrm{H.c.}  \label{Hinh}
\end{equation}

\section{Selective Control in the homogeneous Coupling case}
\subsection{Generalized selectivity}
In order to illustrate how selectivity appears in the $N$-ion
case, let us study the special situation of the $N$ ions coupled
homogeneously to the Raman lasers where $\Omega
_{\mathrm{eff}}^j\equiv\Omega _{\mathrm{eff}} = 2i \eta
_{2}\Omega_{1} \Omega_{2}^{\ast } / \Delta$,
$\Omega_0^j\equiv\Omega_0=2\eta^2_1|\Omega_{1}|^2/\Delta$, and
$\delta_0^j \equiv \delta_0$. In this case, the interaction part
in Hamiltonian~(\ref{HeffN2}) corresponds to an
anti-Tavis-Cummings Model~\cite{Tavis}, a spin $j=N/2$
generalization of the Jaynes-Cummings model~\cite{Jaynes}. In this
case, $\hat{\tilde{J}}^\pm \rightarrow \Omega _{\mathrm{eff}}
\hat{J}^\pm$, and the new collective terms $\hat{J}^\pm$ can be
considered as angular momentum operators, establishing a
permutation symmetry on the ionic subsystem dynamics. That means
that if the system is found at any time inside the symmetric Dicke
subspace~\cite{Dicke54}, associated with  total angular momentum
$j=N/2$, it will stay there along its evolution, reducing the
Hilbert space dimension from $2^N$ to $N+1$. Under this plausible
assumption, the collective operators $ \hat{J}^\pm$ can be
effectively and exclusively rewritten in the symmetric Dicke
subspace via the following assignments
\begin{eqnarray}
\sum_{j=1}^{N}|\mathrm{g}_j\rangle
\langle \mathrm{g}_j| \!\!\!\! && \rightarrow \sum_{k=0}^{N-1}(N-k)|\mathrm{D}_k\rangle
\langle \mathrm{D}_k| , \nonumber \\
\hat{J}^+ \!\!\!\!  &&  \rightarrow \sum_{k=0}^{N-1}
f_k|\mathrm{D}_{k+1}\rangle \langle \mathrm{D}_k| .
\label{assignments}
\end{eqnarray}
Here,
\begin{equation}
|\mathrm{D}_{k}\rangle = \! \bigg(\begin{array}{c}N \\
k \end{array}\bigg)^{-\frac{1}{2}} \!\! \sum_k P_k
(| \mathrm{g}_1,\mathrm{g}_2,...,\mathrm{g}_{N-k},\mathrm{e}_{N-k+1},...,\mathrm{e}_N \rangle)
\end{equation}
are the symmetric Dicke states with $k$
excitations, $\{P_k\}$ is the set of all distinct permutations, and $f_{k}=\sqrt{( k+1)(N-k) }$. It is noteworthy to stress that in the assignments of Eq.~(\ref{assignments}) we have omitted the nonsymmetric components due to the assumed initial symmetric conditions.  In this case, and under homogeneous driving,
we can derive from Eq.~(\ref{HeffN2}) an analog to Eq.~(\ref{Hinh}),
\begin{eqnarray}
\hat{\bar{H}}^{\mathrm{I}}_{\rm{eff}} & \!\! = \!\! & \hbar \hat{a}^{\dagger }\Omega
_{\mathrm{eff} }\sum\limits_{k=0}^{N-1}f_{k} e^{i\Omega _{0}(- \hat{n} +N - 1 - k + \delta_{0})
t}|\mathrm{D}_{k+1}\rangle
\langle \mathrm{D}_k|  \notag \\
&&+\text{\textrm{H.c.}}, \label{Hho}
\end{eqnarray}
a compact expression that will prove useful to study selective
interactions inside the symmetric subspace. Let us consider the
system prepared in the initial state $|N_{0}\rangle
|\mathrm{D}_{k_0}\rangle $. Then, the suitable choice of laser frequencies $\delta_{0}=k_{0}+N_{0}-N+1$ yields a selective resonant coupling inside the subspace $\{ |N_{0}\rangle
|\mathrm{D}_{k_0}\rangle,|N_0+1\rangle |\mathrm{D}_{k_0+1}\rangle
\}$. Moreover, provided that
$\Omega_{0} \gg \Omega_{\mathrm{eff}}$, all other subspaces will remain
off resonance obtaining an effective two-level dynamics. That is,
by selecting a determined subspace the Hamiltonian (\ref{Hho}) can
be written as
\begin{equation}
\hat{\tilde{H}} = \hbar \sqrt {N_0+1} \Omega
_{\mathrm{eff}} f_{k_0}(\hat{\sigma}
_{N_{0}}^{+}\hat{J}^{+}_{k_{0}}+\hat{\sigma} _{N_{0}}^-
\hat{J}^{-}_{k_{0}}) \label{Hn}
\end{equation}
where $\hat{J}^{+}_{k_{0}} = |\mathrm{D}_{k_0+1}\rangle \langle
\mathrm{D}_{k_0}|$ and $\hat{\sigma} _{N_{0}}^{+}=|N_0+1\rangle
\langle N_0|$ are effective spin-1/2 operators stemming from the
reduced Hilbert space of the collective ionic state and the
bosonic field respectively. As we will se below, this selective global interaction will allow us to move confortably inside the symmetric Dicke subspace with high precision~\cite{numerics}.

Considering experimental parameters of ion experiments at NIST
(Boulder)~\cite{Wineland00}, we could achieve an effective
coupling $\Omega_{\mathrm{eff}}\sim 10^5$ Hz, which produces
population inversion in the subspace $\{|N_{0}\rangle
|\mathrm{D}_{k_0}\rangle ,|N_{0}+1\rangle
|\mathrm{D}_{k_0+1}\rangle \} $ in a time $\tau \le 0.1$ ms,
shorter than the typical motional decoherence time $\tau_d \sim
10$ ms.

\subsection{Applications of generalized selectivity}
We discuss now some applications of our method for selectively
manipulating the Dicke subspace. Let us consider the initial state
$|\Psi(0)\rangle = |0\rangle | \mathrm{g...g} \rangle \equiv | 0
\rangle |\mathrm{D}_{0} \rangle$. Tuning into resonance the
subspace transition $\{|0\rangle |\mathrm{D}_{0}\rangle ,|1\rangle
|\mathrm{D}_{1}\rangle \}$, the evolution of this state is given
by
\begin{eqnarray}
|\Psi (t) \rangle &=& \cos ( \sqrt{N} |\Omega _{\mathrm{eff}}| t)
|0\rangle|\mathrm{D}_{0}\rangle \notag \\
 &&-ie^{i\phi }\sin ( \sqrt{ N} |\Omega
_{\mathrm{eff}}| t) |1\rangle |\mathrm{D}_{1}\rangle,
\label{Wt}
\end{eqnarray}
where $\Omega _{\mathrm{eff}}=|\Omega _{\mathrm{eff} }|e^{-i\phi
}$. The one-excitation Dicke state $|\mathrm{D}_{1}\rangle$ is
also a W state
\begin{equation}
|W_{N}\rangle =\frac{1}{\sqrt{N}}(|\mathrm{eg...g}\rangle
+|\mathrm{geg...g}\rangle +...+|\mathrm{g...ge}\rangle ).
\label{WWW}
\end{equation}
This $N$-partite entangled state has great importance in quantum
information theory due to its persistent
entanglement properties, as long as more operational effort is needed to
disentangle this state~\cite{Briegel01}. If this interaction is turned on for a time $2 \sqrt{N} | \Omega
_{\mathrm{eff}}| t=\pi $ and $\phi =\pi /2,$ then Eq.~(\ref{Wt}) becomes
\begin{equation}
|\Psi (t) \rangle =|1\rangle|\mathrm{D}_{1}\rangle \equiv |
1\rangle |W_{N}\rangle, \label{W1}
\end{equation}
yielding state $|W_{N}\rangle $ in the metastable $N$ two-level ions. If the system evolves for a time such that $\cos ( \sqrt{N} |\Omega
_{\mathrm{eff}}| t) =1/\sqrt{N+1}$, then
\begin{equation}
|\Psi ( t) \rangle =|W_{N+1}\rangle,
\end{equation}
where the $( N+1) $-th qubit is the reduced bosonic spin-1/2
system.

Once the system is prepared in the state given in Eq.~(\ref{W1}), and
tuning to resonance the red-sideband subspace transition $\{|1\rangle |\mathrm{D}_{1}\rangle ,|0\rangle
|\mathrm{D}_{2}\rangle \}$, a pulse with Rabi angle $2 \sqrt{N} |\Omega
_{\mathrm{eff}}| t_2=\pi$ will
lead to
\begin{equation}
|\Psi (t_2) \rangle =| 0\rangle |\mathrm{D}_{2}\rangle.
\end{equation}
In this manner, it is clear that a successive application of collective blue- and red-sideband interactions can
produce deterministically and sequentially all symmetric Dicke states $|\mathrm{D}_{k}\rangle$ with number of
excitations $k$.

Another interesting application of multipartite selective
interactions is the possibility to discriminate between ionic
states with different number of excitations. Suppose we have an
ionic state prepared in a superposition of states with different
number of excitations $\sum^{N}_{k=0}c_{k}|\mathrm{D}_{k}\rangle$,
with $\sum^N_{k=0}|c_{k}|^2=1$. For example, this state can
correspond to an atomic coherent state~\cite{Arecchi} given by
$\exp(i\theta \hat{J}_x)|\mathrm{g...g}\rangle$. Note that an
interaction proportional to $\hat{J}_x$ can be generated by
applying a Raman laser field tuned to the carrier transition on
the $N$ ions collectively and homogeneously. The center-of-mass
mode is initialized in the staet $|N_0\rangle$ and we consider an
(additional) ancillary qubit in the ground state
$|\mathrm{g}\rangle _{\mathrm{A}}$. We tune then to resonance the
collective blue-sideband subspace $\{|N_0\rangle
|\mathrm{D}_{k_0-1}\rangle ,|N_0+1\rangle |\mathrm{D}_{k_0}\rangle
\}$, where $|\mathrm{D}_{k_0}\rangle $ is the state with $k_{0}$
excitations we want to discriminate. In this way, after a
collective $\pi$-pulse on the ions, we obtain a state of the form
\begin{eqnarray}
|\Psi
_{1}\rangle
=(c_{k_{0}-1}|N_0+1\rangle |\mathrm{D}_{k_0}\rangle
+|N_0\rangle \sum^{N}_{k\neq
k_{0}-1}c_{k}|\mathrm{D}_{k}\rangle)|\mathrm{g}\rangle
_{\mathrm{A}} . \nonumber \\
\end{eqnarray}
Now, a $\pi $-pulse with the laser field tuned to
the first red sideband on the ancillary qubit leads to
$|\Psi_{2}\rangle=(c_{k_{0}-1}|\mathrm{D}_{k_0}\rangle|\mathrm{e}\rangle
_{\mathrm{A}}+\sum^{N}_{k\neq
k_0-1}c_{k}|\mathrm{D}_{k}\rangle|\mathrm{g}\rangle_{\mathrm{A}})|N_0\rangle$. Then, if we measure the ancilla in the excited
state $|\mathrm{e}\rangle _{\mathrm{A}}$, the collective ionic
state will collapse into the Dicke state $|\mathrm{D}_{k_0}\rangle
$ with $k_{0}$ excitations. Remark that the projection on ancillary state $|\mathrm{e}\rangle _{\mathrm{A}}$, that should happen with a probability $| c_{k_{0}-1} |^2$, can be done with high precision via well established electron-shelving techniques.

On the other hand, it has been shown that the use of selective interactions in a single trapped ion can lead to deterministic and universal
manipulation of the motional state~\cite{Santos05}. Along these lines,
similar manipulation could be implemented here to grant access to arbitrary states inside the symmetric Dicke subspace. In this case, the motional Fock states would be replaced by symmetric states in the internal ionic degrees of freedom with a fixed number of excitations.

\section{Selective Control in the inhomogeneous Coupling case}
In the more general case of ions interacting inhomogeneously with
Raman lasers, we cannot discriminate preselected symmetric Dicke
states. However, multipartite selectivity will still allow us to
manipulate {\it ionic number states}, that is, ionic states with a
determined number of excitations but not necessarily symmetric.
For example, if laser fields interact inhomogeneously with
initially deexcited trapped ions in a carrier-like excitation of
the form $U = \exp{(-i\theta \hat{\tilde{J_x}} )}$, where
$\hat{\tilde{J_x}} = \hat{\tilde{J}}^{+} + \hat{\tilde{J}}^{-}$,
this will not lead to a superposition of symmetric Dicke states.
On the opposite, this will lead to a superposition of nonsymmetric
collective number states arising from the action of the operators
$\hat{\tilde{J}}^{+}$ and $\hat{\tilde{J}}^{-}$ on the collective
ionic states. It is known that to deal with the unitary evolution
of high-dimensional inhomogeneously coupled systems is  extremely
difficult~\cite{Lopez06,Lopez07}. In this case, instead of writing
the Hamiltonian~(\ref{HeffN2}) in the basis of the symmetric Dicke
states, as in Eq.~(\ref{Hho}), we should write it in the
corresponding basis of nonsymmetric collective number states
$|\mathrm{\tilde{D}^{\ell}}_{k}\rangle$ with $k$ excitations. In
this way, we may look for conditions to set into resonance a
determined subspace. States
$|\mathrm{\tilde{D}^{\ell}}_{k}\rangle$ appear naturally from
successive applications of $\hat{\tilde{J}}^{+}$ and
$\hat{\tilde{J}}^{-}$ on a given initial collective state. The
index $\ell$ accounts for the fact that, depending on the number
of ionic excitations, there could exist more than one nonsymmetric
collective state with a determined number of excitations. In the
same spirit of Eq.~(\ref{HeffN}) , we can write the associated
Hamiltonian

\begin{eqnarray}
\hat{\bar{H}}^{\mathrm{I}}_{\rm{eff}} &=&-\hbar
\sum_{k,\ell}\langle\mathrm{\tilde{D}^{\ell}}_{k}
|\hat{\Delta} |\mathrm{\tilde{D}^{\ell}}_{k}\rangle
|\mathrm{\tilde{D}^{\ell}}_{k}\rangle\langle\mathrm{\tilde{D}^{\ell}}_{k}| \notag \\
&& + \hbar \hat{a}^{\dagger}\sum_{k,\ell}\tilde{\Omega}_{\mathrm{eff}}^{k,\ell}
|\mathrm{\tilde{D}^{\ell}}_{k+1}\rangle
\langle\mathrm{\tilde{D}^{\ell}}_{k}|+\text{\textrm{H.c.}}
\label{Hinh2}
\end{eqnarray}
Here, $\tilde{\Omega}_{\mathrm{eff}}^{\ell,k}$ is the
new effective coupling constant, which in the homogeneous case
corresponds to $\Omega_{\mathrm{eff}}$. As in the homogeneous
case, if $\Omega_0 \gg \tilde{\Omega}_{\mathrm{eff}}^{\ell,k}$, we
can tune to resonance a determined subspace, for example the inhomogeneous blue-sideband doublet
$\{|N_0\rangle |\mathrm{\tilde{D}^{\ell}}_{k_0}\rangle
,|N_0+1\rangle |\mathrm{\tilde{D}^{\ell}}_{k_0+1}\rangle
\}$. In this case, from Hamiltonian~(\ref{Hinh2}) in the
interaction picture, we can derive that the condition to tune to
resonance this subspace is
$\langle\mathrm{\tilde{D}^{\ell}}_{k_0+1}
|\hat{\Delta}_{N_0+1}|\mathrm{\tilde{D}^{\ell}}_{k_0+1}\rangle-\langle\mathrm{\tilde{D}^{\ell}}_{k_0}
|\hat{\Delta}_{N_0}|\mathrm{\tilde{D}^{\ell}}_{k_0}\rangle=0$. This
condition can be fulfilled by compensating the detuning
$\hat{\Delta}$ through shifts in the lasers frequencies for fixed
values of $\delta_0^j$, depending on the subspace we want to
select. This procedure is similar to the homogeneous case, but now
$\delta_0^j$ will be inhomogenously distributed, that is, different for each ion.

\section{Conclusions}

In conclusion, we have introduced a selective technique that allow
a collective manipulation of the ionic degrees of freedom inside
the symmetric Dicke subspace. We have studied the homogeneous and
inhomogeneous cases, showing applications in both cases, mainly
related to the generation and control of number states in the
ionic external and internal degrees of freedom. We believe that
the introduced concepts may inspire similar physics in other
quantum-optical setups with diverse applications, and that they
might even be helpful to transfer collective atomic number states
to propagating fields.

\vspace*{0cm}

\section*{ACKNOWLEDGMENTS}

CEL is financially supported by MECESUP USA0108 and CONICYT, JCR
by Fondecyt 1070157 and Milenio ICM P02-049F, and ES by DFG SFB
631, EU EuroSQIP projects, and the German Excellence Initiative via
the ``Nanosystems Initiative Munich (NIM)''. CEL also thanks to
DIGEGRA USACH and Jan von Delft for hospitality at Ludwig-Maximilian University (Munich).

\end{document}